\documentclass[utf8]{frontiersSCNS}

\usepackage{url,hyperref,lineno,microtype,subcaption}
\usepackage[onehalfspacing]{setspace}

\def\keyFont{\fontsize{8}{11}\helveticabold }
\def\firstAuthorLast{La Mura {et~al.}} 
\def\Authors{Giovanni La Mura\,$^{1,*}$, Marco Berton\,$^{1,2}$, Sina Chen\,$^1$, Stefano Ciroi\,$^1$, Enrico Congiu\,$^{1,2}$, Valentina Cracco\,$^1$, Michele Frezzato\,$^1$, Piero Rafanelli\,$^1$}
\def\Address{$^{1}$Department of Physics and Astronomy, University of Padua, Padua, Italy \\
  $^{2}$Astronomical Observatory of Brera, National Institute for Astrophysics (INAF), Milan, Italy}
\def\corrAuthor{Giovanni La Mura, Vicolo dell'Osservatorio 3, I-35122 Padua}

\def\corrEmail{giovanni.lamura@unipd.it}

\begin{document}
\onecolumn
\firstpage{1}

\title[AGN multi-frequency databases]{Multi-frequency databases for AGN investigation - Results and perspectives}

\author[\firstAuthorLast ]{\Authors} 
\address{} 
\correspondance{} 

\extraAuth{}

\maketitle

\begin{abstract}

\section{}
Active Galactic Nuclei (AGNs) are characterized by emission of radiation over more than 10 orders of magnitude in frequency. Therefore, the execution of extensive surveys of the sky, with different types of detectors, is providing the attractive possibility to identify and to investigate the properties of AGNs on very large statistical samples. Thanks to the large spectroscopic surveys that allow detailed investigation of many of these sources, we have the opportunity to place new constraints on the nature and evolution of AGNs and to investigate their relations with the host systems. In this contribution we present the results that can be obtained by using a new interactive catalogue that we developed to investigate the range of AGN spectral energy distributions (SEDs). We present simple SED models based on data collected in the catalogue and discuss their relations with optical spectra obtained by follow up observations. We compare our findings with the expectations based on the AGN Unification Model, and we discuss the perspectives of multi-wavelength approaches to address AGN related processes such as black hole accretion and acceleration of relativistic jets.

\tiny
\keyFont{ \section{Keywords:} galaxies: active, galaxies: nuclei, quasars: emission lines, quasars: supermassive black holes, galaxies: jets, catalogs, surveys} 
\end{abstract}

\section{Introduction}

Since their first identification as sources of prominent emission lines from ionized gas in the core of galaxies with exceptionally bright centers \citep{Seyfert43}, {\it Active Galactic Nuclei} (AGNs) became a primary source of discoveries and new challenges in Astrophysics. In addition to being the most energetic non-transient objects that we currently know, maintaining luminosities of the order of $L > 10^{41}\, \mathrm{erg\, s}^{-1}$ for timescales spanning from several thousands up to hundred millions years, they populate both the local and the remote Universe, suggesting that their existence must have a relevant impact on the evolution of cosmic structures. The well known paradigm that identifies their energy source as the conversion of gravitational binding energy of material accreted by a Super Massive Black Hole (SMBH) into radiative power accounts very well for most of their observational properties \citep{Blandford79,Blandford86} and it is largely supported by the strong evidence for the existence of SMBHs in the nuclei of all massive galaxies \citep{Magorrian98,Ferrarese00}. It is well established that the process of matter accretion onto SMBHs can give raise to different mechanisms resulting in the production of a complex spectrum of electro-magnetic radiation that, depending on the characteristics of the source, may exhibit various emission and absorption components. Attempts to explain the resulting wealth of observational features in a single physical picture led to the formulation of the \emph{AGN Unification Model} \citep{Antonucci93,Urry95}.

In spite of major advances in our understanding of the physics of AGNs, today we still seek the answer of some fundamental questions, such as how nuclear activity is triggered, what role does it play on the evolution of galaxies and which are the most relevant physical processes that govern the mass flow in the close vicinity of the SMBH, controlling the formation of structures like accretion disks and relativistic jets. While waiting for the deployment of instrumental facilities that might directly address part of these questions, such as the \emph{James Webb Space Telescope} (JWST), the \emph{Square Kilometer Array} (SKA), or the \emph{European Extremely Large Telescope} (E-ELT), among others, the approach of testing the properties of AGNs taking advantage of the complexity and the extension of their spectral energy distribution (SED) on large statistical populations is providing very attractive results \citep{Massaro11,dabrusco12}.

With this contribution, it is our aim to present an innovative approach to collect multi-frequency information from astronomical databases, in order to identify different types of AGNs. The gathered information can be used in connection with spectroscopic surveys, monitoring experiments and follow-up observations in order to validate the predictions of AGN physical models and to clarify some of the issues that are currently under debate. In the following sections we shall therefore present our work on AGN databases (\S2), discuss its comparison with other existing services (\S3), and finally provide considerations on the results that have so far been achieved, with a mention on their possible development.

\section{An interactive AGN multi-frequency observation archive}

By identifying AGNs with objects where the central SMBH of a galaxy is persistently fed by an accretion flow, it can be roughly estimated that, from the highest power Quasi Stellar Object (QSO) to the much dimmer Low-Ionization Nuclear Emission Region (LINER), approximately 10\% of all galaxies host some kind of nuclear activity. The task to collect into a single catalog the most fundamental information, such as position, distance, apparent magnitude and luminosity estimates, becomes itself a non trivial problem with the resulting numbers. The \emph{Catalog of Quasars and Active Galactic Nuclei} \citep{VeronCetty10}, in its $13^{\rm th}$ edition, represented a major attempt to assemble a systematic list of active galaxies, providing basic information. With its 168480 main entries, however, it represents only a small fraction of the millions potentially detectable sources, as demonstrated by \citet{Flesch15}. In addition to designing an effective scheme to handle the catalogue contents, the large number of entries implies a coarse classification of objects with a non-negligible chance to overlook important features or to fail in spotting unreliable measurements.

It is now clear that in the case of AGNs a proper and physically meaningful classification is fundamental, both for our current understanding, as well as for its development perspectives. The execution of large survey and monitoring programs with publicly available spectroscopic observations, such as the Sloan Digital Sky Survey \citep[SDSS]{York00,Alam15} and the 6dF Galaxy Redshift Survey \citep[6dFGS]{Jones04,Jones09}, combined with the deployment of instruments able to explore the electromagnetic radiation with unprecedented sensitivity and resolution, especially in the energies of the X rays and above, now provide the possibility to probe the physics of many objects in very high detail. Taking advantage of the increase in the computational power and in the efficiency of data transmission networks available to final users, which characterized the technological progress of the last years, we explored the possibility to develop a new strategy to collect multi-frequency data and to use them for the investigation of AGNs. Due to the complex mechanisms that are involved in the emission of their radiation, generally resulting from the combination of both thermal and non-thermal processes, the SEDs of AGNs often span several orders of magnitude in frequency, effectively distinguishing these objects from quiescent and star-forming galaxies, whose emission is mostly confined in the IR-optical-UV range. The use of this property, however, is not always the best option to search for AGNs, since intrinsic or external absorption processes can lead to the loss of important frequency ranges. Limiting the analysis to very specific radiation bands, on the other hand, may lead to substantial selection effects and, therefore, restrict the scope of investigations to under-representative samples. Some additional methods to detect AGNs rely on variability investigations, but again they are prone to the observability constraints of the selected frequency and they can be considerably expensive, in terms of observational efforts, due to the necessity to monitor many targets on a regular schedule.

In order to overcome, at least in part, the difficulties listed above, we explored the possibility of building an interactive AGN observation catalogue, named AGN Multi-Wavelength Catalogue (AGNMWC), that users can tune with minimal effort to match their specific requirements. Because our aim has been to set up an instrument that can manage large sets of information, without losing the possibility to explore single objects in detail, we adopted a structured database architecture, as our strategy to store data, and we developed a client application, written in Java language, to grant compliance with any operating system hosting a Java Runtime Environment (JRE). The innovative aspect of this solution is that it is not constrained to access a specific data source, which might become outdated or unavailable, but it allows the user to build customized catalogues, using data from virtually any source, provided that the correct information on how to manage the input is given. The adopted strategy is to connect to an SQL database, presently managed either by \emph{MySQL} or \emph{SQL Server}, that must be available on the user's workstation or through network or Internet connection. The Java application is therefore used to import data selected by the user and to build a multi-frequency catalogue in the database where every data source is associated with a set of customizable options that define what type of information is being stored. The application also connects through the Internet to the SDSS and the 6dF spectroscopic databases, to visualize spectral previews, if available, and to the Digitized Sky Survey, to show maps of the sources in the optical $R$ band.

\section{Comparison with existing catalogues and on-line services}

One of the key features of scientific investigation is the repeatability of experiences. In the case of Astrophysics this requirement can hardly be matched, because many observations, especially those obtained by space-born detectors, are prone to a continuous deterioration of the instrument. Moreover, the study of variable sources, like many processes connected with AGN activity, is often related to unpredictable situations, which rarely repeat under similar conditions. For these reasons, the possibility to repeat experiences is generally replaced with an accurate knowledge of the observation target and of the conditions of instruments at the time of the observation. Holding such information is the main goal of the data archives maintained by most observatories and experiment collaborations. In general, each instrument has its own characteristics, which require different calibrating information and reduction approaches, in order to extract the final data products. Replicating such information to meet the requirements of different users, or communicating it upon request, usually leads to large costs in terms of data storage or transmission efficiency.

In recent times, thanks especially to a fast increase in the efficiency of data transmission networks, several tools have been developed, which connect different types of archives. These tools lead to the collection of high-level data products, offering selective and customizable output. Very well known examples are the Nasa Extragalactic Database (NED)\footnote{\texttt{https://ned.ipac.caltech.edu/}} and the Vizier service\footnote{\texttt{http://vizier.u-strasbg.fr/viz-bin/VizieR}}, which, connecting different catalogues and data sources are able to plot SEDs and to connect to external spectroscopic servers (chiefly the SDSS and the 6dF databases). More recently, the \emph{Virtual Observatory} project\footnote{\texttt{http://www.ivoa.net/index.html}} attempted the creation of a more interactive environment, providing tools to inspect raw and calibrated data from various archives. Other mission specific tools to analyze data at various levels are provided, for instance, by the High Energy Astrophysic Science Archive Center (HEASARC)\footnote{\texttt{https://heasarc.gsfc.nasa.gov/}} and the Italian Space Agency Science Data Center (ASDC)\footnote{\texttt{https://tools.asdc.asi.it/}}, which additionally proposes powerful tools to fit analytical models to specific types of SEDs. The possibility to build a catalogue of multi-frequency AGN observations based on the contemporary availability of data has also been thoroughly explored \citep{Padovani97}.

All of the aforementioned solutions are particularly useful either to perform detailed investigations of single targets or to quickly manage large data samples, but they often cannot work as effectively in both modes, or provide a customizeable and easy to update solution. The choice to set up a new archive interface, therefore, descends from the importance of collecting data from the large archives of AGN observations that we currently possess, while maintaining the ability to develop customizable access to advanced information, such as spectroscopy and time resolution. All such features should be combined with the ability to perform statistical studies of SEDs, fitted through combinations of thermal and non-thermal radiation sources.

\section{Results and perspectives}

Due to the complex combination of thermal and non-thermal contributions that participate to their spectrum, involving black-body radiation from the accretion disk, synchrotron emission from relativistic plasmas and inverse Compton scattering of seed photons in the disk corona and possibly in jets, AGNs are commonly detected over a broad range of frequencies. As a result AGNs are often characterized by a high-energy radiation excess and X-ray surveys are very effective to identify them. Moreover they are the dominant population of extragalactic sources in the range of $\gamma$-ray emission \citep{Acero15}. Taking advantage of the availability of spectroscopic surveys, we are able to investigate the SED properties of multi-frequency emitters, in connection with the optical spectroscopic properties that lead us to identify the existence of a specific type of nuclear activity. Fig.~1 illustrates an example application of AGNMWC to set up such kind of investigation. Starting from the $2^{\rm nd}$ ROSAT all sky survey catalogue \citep[2RXS,][]{Boller16} and requiring coincident detections in the WISE, GALEX and USNO \citep{Monet03} at an angular separation of $r < 15\,$arcsec (to account for the limited spatial resolution of ROSAT), we select 53683 sources with multi-frequency detection, distributed at high galactic latitude. Using AGNMWC to import observations obtained from the data sources listed in Table~1, we are able to reconstruct the SED of targets and to model their features in terms of power-law and black-body components. By comparing the models of different types of sources, we can immediately appreciate how the transition from Type 1 to Type 2 objects occurs in a consistent way with the expected effects of an increasing amount of obscuration towards the central source. The case of 3C~273, a prototypical Type 1 QSO belonging to the Flat Spectrum Radio Quasar (FSRQ) blazar family, is a fair example of AGN SED with prominent non-thermal emission and a clear signature of a thermal excess, having an estimated temperature of $2.58 \cdot 10^{4}\,$K, in very good agreement with the expectations for an AGN seen without significant obscuration towards the central engine. Conversely, {\bf in} the SED of NGC~1068, a Type 2 source that has well established evidence for a hidden Broad Line Region \citep[BLR,][]{Antonucci85}, the non-thermal contribution is energetically overwhelmed by the thermal emission of an optically thick obscuring structure that absorbs most of the ionizing radiation of the central source, with just the exception of the highest energy, more penetrating photons, and radiates it back isotropically, in the form of an IR bump. The search of similar features in the SEDs of objects detected at various frequencies leads to the identification of AGN candidates located at high Galactic latitudes all over the sky. The distribution of such objects strongly argues in favor of their extragalactic origin and the AGN nature can be directly assessed, if the source lies in the footprint of a spectroscopic survey or it can be associated to an optical counterpart for spectroscopic follow-up. Looking at the sources located in the footprint of the SDSS-DR 12, we find 6127 objects with spectra, out of which 5901 are spectroscopically confirmed AGNs, while the remaining are mainly interacting star systems. By requiring that the X-ray spectra of the sources are dominated by a power-law contribution, we reduce to a total number of 2389 sources, 2300 out of which are confirmed active galaxies. Similar methods have been widely used in the literature \citep[e.g][]{Massaro14, Massaro15, Alvarez16a, Alvarez16b, Alvarez16c}, and they were specifically adapted in our catalogues to investigate the nature of unclassified $\gamma$-ray sources \citep{Chiaro16, LaMura17}.

\begin{table}[t]
\begin{center}
\caption{Data sources for the selection of multi-wavelength SED points. \label{table:tab01}}
\begin{tabular}{cccc}
\hline
\hline
{\bf Instr. / Catalogue} & $\log \nu$ (Hz) & {\bf Band} & {\bf Reference} \\
\hline
SUMSS & $8.93$ & Radio & \citet{Mauch03} \\
NVSS & $9.15$ & Radio & \citet{Condon98} \\
IRAS & $12.48 -  13.40$ & FIR & \citet{Helou88} \\
WISE & $13.13 - 13.94$ & FIR & \citet{Cutri12} \\
2MASS & $14.14 - 14.38$ & NIR & \citet{Skrutskie06} \\
GALEX & $15.00 - 15.30$ & UV & \citet{Bianchi11} \\
ROSAT & $16.86 - 17.23$ & Soft-X & \citet{Boller16} \\
XMM-Newton & $16.86 - 18.16$ & X rays & \citet{Rosen16} \\
INTEGRAL & $18.70 - 19.00$ & Soft $\gamma$\ rays & \citet{Bird10} \\
Fermi/LAT & $23.00 - 26.00$ & $\gamma$\ rays & \citet{Acero15} \\
\hline
\end{tabular}
\end{center}
\end{table}
An additional advantage of developing a customizable service to investigate the multi-wavelength properties of AGNs stems from the possibility to explore time domain properties. Indeed, one of the most striking properties of AGNs, particularly in the case of Type 1 sources and especially for blazars, is a strong and mostly irregular variability. The origin of this behavior is related to the details of the accretion process in the central engine and the influence of changes in one spectral component on the others may lead to unveil the structure of the source beyond the current limits of observational resolution \citep{Blandford82, Peterson98}. An example of such application is illustrated in Fig.~2, for the case of the FSRQ blazar 3C~345 \citep{Berton17}. During the period monitored by the Fermi/LAT $\gamma$-ray telescope between August 2008 and October 2016, this source showed a significant decrease of high-energy activity, which, in spite of the occurrence of strong outbursts, apparently involved its overall emission. Like many other blazars, its SED is energetically dominated by the non-thermal processes arising in the jet and the observed decrease in luminosity might be interpreted as a loss of jet power. The ejection of new relativistic plasma blobs, reported by VLBI radio observations in correspondence with some flares represents episodic breaks in this general trend, which, in some cases, appears to affect other spectral components, like the prominent broad Mg~{\small II} $\lambda 2798$ emission line. This line, which in general shows a well defined anti-correlation between equivalent width and continuum intensity \citep[i.e. a Baldwin Effect,][]{Baldwin77, Pogge92}, exhibited a strong variation shortly after a $\gamma$-ray flare without any evidence of changes in the underlying continuum, suggesting that the flare occurred very close to the central SMBH and within the BLR. A subsequent analysis of its spectrum and SED shows that, while moving from the maximum activity to the minimum observed during the monitoring, the source featured a decrease of jet emission power, but a relevant enhancement in the relative importance of the thermal component, for which we also observe a slight increase of temperature. A possible explanation of such a behavior may reside in the fact that the processes, which feed power to the jet, are likely slowing the accretion rate. In such circumstances, it could be possible that the ejection of a plasma blob from very close to the black hole affects the coupling of jet and accretion disk, favoring the accretion flow with respect to the jet immediately after the outburst. Clearly much more investigation of similar processes in this and other sources is required to verify this possibility. A summary of the models that were used to reconstruct the illustrated SEDs is reported in Table~2.

\begin{table}[t]
  \caption{SED fitting models.}
  \centering
  \begin{tabular}{lccccccc}
    \hline
    \hline
      {\bf Object} & {\bf Function} & $\log \nu_{min}^{(a)}$ & $\log \nu_{max}^{(a)}$ & {\bf Norm.} & {\bf Index}$^{(b)}$ & {\bf Temp.} & $\chi_{red}^2$ \\
      \hline
      3C 273   & Power law  & $8.6$ & $13.1$ & $2.30 \cdot 10^{-18} $ & $0.60$ & -- & \\
               & Power law  & $13.1$ & $17.6$ & $1.31 \cdot 10^{-6} $ & $-0.31$ & -- & \\
               & Power law  & $17.1$ & $22.6$ & $1.26 \cdot 10^{-23}$ & $0.70$ & -- & \\
               & Black body & -- & -- & $1.03 \cdot 10^{-12}$ & -- & $25800\,$K & $1.11$ \\
      \hline
      NGC 1068 & Power law  & $8.5$ & $13.0$ & $2.49 \cdot 10^{-28}$ & $1.52$ & -- & \\
               & Power law  & $12.8$ & $17.5$ & $7.068 \cdot 10^2$ & $-0.85$ & -- & \\
               & Power law  & $17.0$ & $24.0$ & $4.82 \cdot 10^{-13}$ & $0.05$ & -- & \\
               & Black body & -- & -- & $1.04 \cdot 10^{-4}$ & -- & $540\,$K & $1.17$ \\
      \hline
      3C 345 (high) & Power law  & $11.0$ & $16.8$ & $5.193 \cdot 10^{-10}$ & $-0.19$ & -- & \\
               & Power law  & $17.0$ & $24.5$ & $8.28 \cdot 10^{-19}$ & $0.31$ & -- & \\
               & Black body & -- & -- & $7.51 \cdot 10^{-17}$ & -- & $31340\,$K & $1.05$ \\
      \hline
      3C 345 (low) & Power law  & $10.1$ & $17.0$ & $1.096 \cdot 10^{-12}$ & $-0.06$ & -- & \\
               & Power law  & $17.0$ & $24.5$ & $7.07 \cdot 10^{-18}$ & $0.24$ & -- & \\
               & Black body & -- & -- & $3.38 \cdot 10^{-17}$ & -- & $38630\,$K & $1.06$ \\
      \hline
  \end{tabular}
  \flushleft{\footnotesize $^{(a)}$ Logarithm of exponential cut-off frequency given in Hz. \\
  $^{(b)}$ The power-law index is given according to the notation $\nu F_\nu \propto \nu^\alpha$.}
\end{table}
In order to take full advantage of the opportunity to investigate AGN SEDs with the wealth of available observing material, we are currently developing our interactive catalogue tools in the framework of possible connections with an ever increasing number of online services and particularly matching the procedural strategies of the Virtual Observatory. Although several opportunities to improve AGNMWC can still be pursued, especially with the inclusion of more detailed physical models to fit the target SEDs and the introduction of more suitable stand-alone solutions that may rely on lighter requirements on the connection with SQL driven databases, we maintain a test suite of the tool at the URL address: \begin{center}
\texttt{https://1drv.ms/u/s!AngiMxTxSRoxgjnYsXm3-DuHub2u}
\end{center}
that is used for testing and development purposes. The online package includes the Java client application, the example catalogue described in this study and a technical description illustating the procedures to build a compliant database. Its system requirements are limited to the presence of a JRE v1.6 or above and to the possibility to connect to an SQL database managed either by MySQL v5.6.19 or by MS SQL Server 2008, or following releases.

\section*{Conflict of Interest Statement}

The authors declare that the research was conducted in the absence of any commercial or financial relationships that could be construed as a potential conflict of interest.

\section*{Author Contributions}

GLM is the developer of the AGN Multi-Wavelength Catalogue, PI of AGN selection procedures in databases and author of the manuscript's text. MB is PI of the AGN monitoring programs that provided data on 3C~345. SCh. contributed to source association and spectroscopic follow-up in the Southern Hemisphere, SC contributed as data reduction specialist, EC, VC, MF contributed multi-wavelength data analysis, target selection and SED modeling, while PR is the concept creator for the Multi-Wavelength Catalogue and the scientific supervisor of the project. All authors contributed to text revision and improvement.

\section*{Funding}
Funding for this research program has been provided by the Department of Physics and Astronomy of the University of Padua and by the European Space Agency (ESA), under Express Procurement Contract No. 4000111138/14/NL/CB/gp. Additional support from ASTROMUNDUS mobility grants is also gratefully acknowledged.

\section*{Acknowledgments}

This research has made use of the NASA/IPAC Extragalactic Database (NED) which is operated by the Jet Propulsion Laboratory, California Institute of Technology, under contract with the National Aeronautics and Space Administration.

Funding for the Sloan Digital Sky Survey IV has been provided by the Alfred P. Sloan Foundation, the U.S. Department of Energy Office of Science, and the Participating Institutions. SDSS acknowledges support and resources from the Center for High-Performance Computing at the University of Utah. The SDSS web site is www.sdss.org.

SDSS is managed by the Astrophysical Research Consortium for the Participating Institutions of the SDSS Collaboration including the Brazilian Participation Group, the Carnegie Institution for Science, Carnegie Mellon University, the Chilean Participation Group, the French Participation Group, Harvard-Smithsonian Center for Astrophysics, Instituto de Astrof\'{i}sica de Canarias, The Johns Hopkins University, Kavli Institute for the Physics and Mathematics of the Universe (IPMU) / University of Tokyo, Lawrence Berkeley National Laboratory, Leibniz Institut f\"ur Astrophysik Potsdam (AIP), Max-Planck-Institut f\"ur Astronomie (MPIA Heidelberg), Max-Planck-Institut f\"ur Astrophysik (MPA Garching), Max-Planck-Institut f\"ur Extraterrestrische Physik (MPE), National Astronomical Observatories of China, New Mexico State University, New York University, University of Notre Dame, Observat\'orio Nacional / MCTI, The Ohio State University, Pennsylvania State University, Shanghai Astronomical Observatory, United Kingdom Participation Group, Universidad Nacional Aut\'onoma de M\'exico, University of Arizona, University of Colorado Boulder, University of Oxford, University of Portsmouth, University of Utah, University of Virginia, University of Washington, University of Wisconsin, Vanderbilt University, and Yale University.

This work is based on observations collected at Copernico telescope (Asiago, Italy) of the INAF - Osservatorio Astronomico di Padova and on observations collected with the 1.22m \textit{Galileo} telescope of the Asiago Astrophysical Observatory, operated by the Department of Physics and Astronomy "G. Galilei" of the University of Padova.


\newcommand{\aap}{A\&A}
\newcommand{\aj}{AJ}
\newcommand{\apj}{ApJ}
\newcommand{\apjl}{ApJL}
\newcommand{\apjs}{ApJS}
\newcommand{\apss}{Ap\&SS}
\newcommand{\araa}{ARA\&A}
\newcommand{\memsai}{MemSAIt}
\newcommand{\mnras}{MNRAS}
\newcommand{\pasa}{PASA}
\newcommand{\pasp}{PASP}
\bibliographystyle{frontiersinSCNS_ENG_HUMS} 
\bibliography{lamura_v02}

\section*{Figure captions}


\begin{figure}[h!]
\begin{center}
\includegraphics[width=0.35\textwidth]{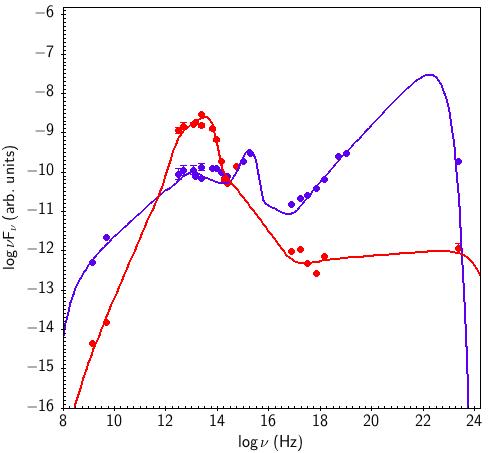}
\includegraphics[width=0.55\textwidth]{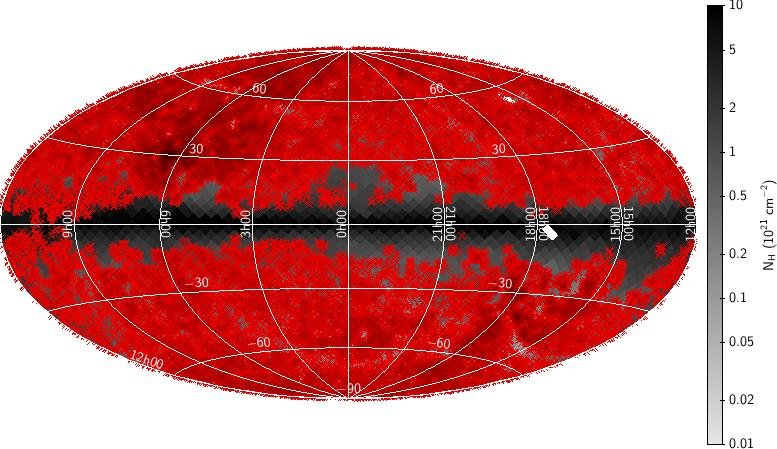}
\end{center}
\caption{{\bf Left:} Examples of SED models for proto-typical Type 1 (3C~273, blue points and continuous curve) and Type 2 (NGC~1068, red points and continuous curve) obtained from the cross-match of NVSS, WISE, IRAS, 2MASS, USNO B2, GALEX, XMM-Newton, INTEGRAL and Fermi/LAT data. The SEDs have been modeled through the combination of three exponential cut-off power laws and a black-body contribution and normalized to the same optical flux. We note the energetically dominant non-thermal emission with a hot UV thermal contribution in the case of the Type 1 SED, compared to the IR thermal excess due to radiation reprocessing in the warm obscuring medium of the Type 2 source. {\bf Right:} Aitoff projected distribution of 53683 sources, selected on the basis of detected emission over the frequency range spanning from IR to X rays, plotted in galactic coordinates over a gray-scale map of H{\small I} column density in the Milky Way. Absorption of radiation and source confusion are still the main issues affecting the detection of extragalactic objects close to the Galactic plane. \label{fig:1}}
\end{figure}

\begin{figure}[h!]
\begin{center}
\includegraphics[width=0.45\textwidth]{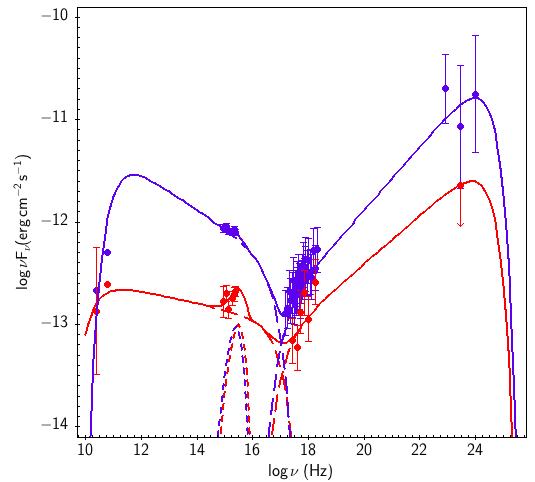}
\includegraphics[width=0.45\textwidth]{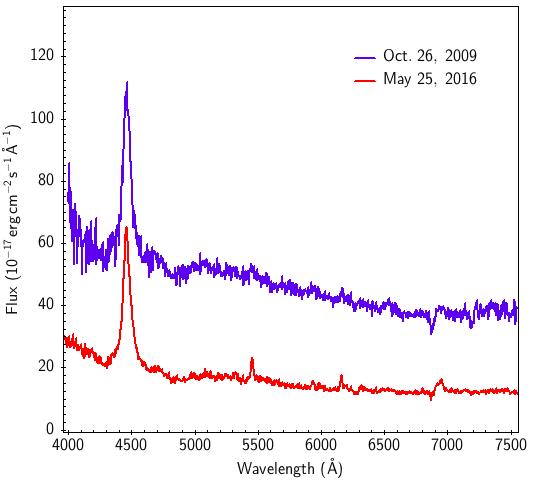}
\end{center}
\caption{SED fits (left panel) and corresponding optical spectra (right panel) of the FSRQ blazar 3C~345 observed during a high state (October 26, 2009; blue points and lines) and a low state (May 25, 2016; red points and line). Non-thermal power laws have been represented as long-dashed lines, while short-dashed curves are thermal contributions. Solid lines are used to show total models. The power-law dominance for this object suggests that the main source of radiation probably arises from the relativistically beamed contribution of the jet, as expected for blazars. In the low activity state, however, a slight but sensible increase in the relative importance of the thermal component can be appreciated, suggesting that a weaker jet activity is likely being associated with an increased accretion efficiency and a higher temperature of the accretion disk. \label{fig:2}}
\end{figure}





\end{document}